\documentclass[conference]{IEEEtran}
\usepackage{cite}
\usepackage{amsmath,amssymb,amsfonts}
\usepackage[ruled,linesnumbered]{algorithm2e}
\SetKwInput{Require}{Require}
\SetKwInput{Ensure}{Ensure}
\usepackage{graphicx}
\usepackage{textcomp}
\usepackage{xcolor}
\usepackage{hyperref}
\usepackage[short]{optidef}
\usepackage{csvsimple,booktabs}
\usepackage[noend]{algpseudocode}

\usepackage[draft, commentmarkup=todo,
  todonotes={textsize=tiny, textwidth=0.83in}]{changes}
\definechangesauthor[name={Ilya Safro}, color=yellow]{IS}

\definechangesauthor[name=jl, color=orange]{jl}

\def\BibTeX{{\rm B\kern-.05em{\sc i\kern-.025em b}\kern-.08em
    T\kern-.1667em\lower.7ex\hbox{E}\kern-.125emX}}
\begin{document}

\title{Decomposition Based Refinement for the Network Interdiction Problem 
%\\
}

\author{\IEEEauthorblockN{Krish Matta}
\IEEEauthorblockA{\textit{School of Computer Science} \\
\textit{Carnegie Mellon University}\\
Pittsburgh, USA \\
kmatta@andrew.cmu.edu}
\and
\IEEEauthorblockN{Xiaoyuan Liu}
\IEEEauthorblockA{\textit{Fujitsu Research of America} \\
Sunnyvale, USA \\
xliu@fujitsu.com}
\and
\IEEEauthorblockN{Ilya Safro}
\IEEEauthorblockA{\textit{Department of Computer and Information Sciences} \\
\textit{University of Delaware}\\
Newark, USA \\
isafro@udel.edu}
}

\maketitle

\begin{abstract}
The shortest path network interdiction (SPNI) problem poses significant computational challenges due to its NP-hardness. Current solutions, primarily based on integer programming methods, are inefficient for large-scale instances. In this paper, we introduce a novel hybrid algorithm that can utilize Ising Processing Units (IPUs) alongside classical solvers. This approach decomposes the problem into manageable sub-problems, which are then offloaded to the slow but high-quality classical solvers or IPU. Results are subsequently recombined to form a global solution. Our method demonstrates comparable quality to existing whole problem solvers while reducing computational time for large-scale instances. Furthermore, our approach is amenable to parallelization, allowing for simultaneous processing of decomposed sub-problems.\\ Reproducibility: Our source code and experimental results are available at \url{https://github.com/krishxmatta/network-interdiction}. 
\end{abstract}


\section{Introduction}
Network interdiction  refers to a class of challenging combinatorial optimization problems that involve strategically disrupting flows in a network to achieve specific objectives \cite{smith2020survey}. These problems can be characterized as a game between attackers and defenders of a network. The defender seeks to optimize a predefined objective across the network, such as maximizing flow between two nodes, while the attacker aims to impede the defender's objective by inflicting maximum disruption upon the network. This disruption may manifest in a variety of forms, such as targeted attacks on the network arcs aiming to destroy or impair them.

The topic of network interdiction has gained significant attention due to its ability to model complex situations in the defense domain. For example, Ghare et al. \cite{https://doi.org/10.1002/nav.3800180103} demonstrated how network interdiction can be used in wartime to capacitate an enemy's supply network to maximally disrupt the flow of enemy troops. Network interdiction can additionally be used to identify weaknesses in critical infrastructure to make them more resilient to terrorist attacks and natural disasters. An example of such an application can be found in a study by Salmerón et al. \cite{4762170}, which details the use of network interdiction to model attacks on large-scale power grids. The study then utilizes this model to determine how best to minimize economic loss resulting from these attacks. Another study by Nandi et al. \cite{NANDI2016118} details the usage of a network interdiction model on cyber attack graphs to pinpoint vulnerabilities in cyberinfrastructures, thereby helping prevent organizations from cyberattacks. Such problems are even extremely useful outside of military applications, e.g. in this line of work  \cite{leyffer2013fast,altunay2011optimal,goldberg2015optimal} the authors show how network interdiction can be used to minimize the generally defined infection spread and attacks.

Within this paper, we study the Shortest Path Network Interdiction (SPNI) problem. In this problem, the defender wishes to traverse the minimum length path between two specified nodes $s$ and $t$ in a directed network. The attacker uses their limited resources to destroy certain arcs, or increase their effective length, to increase the defender's shortest $s$-$t$ path length. Each arc is either interdicted or not, and additionally, it is known beforehand how much an arc costs to interdict and how much interdiction of an arc causes its effective length to increase. When solving SPNI, we take the viewpoint of the attacker, and thus our goal is to maximize the shortest $s$-$t$ path length.

\textbf{Our contribution} The SPNI is an $\mathcal{NP}$-Hard problem \cite{https://doi.org/10.1002/net.10039}. %, no optimal solution can be found deterministically in a polynomial amount of time, meaning that 
To the best of our knowledge, almost all current SPNI algorithms rely on integer programming methods. On large problem instances, these methods are time costly---for example, in one of the most important works regarding SPNI, the largest network solved by Israeli et al. had 240 nodes and 1,042 arcs \cite{https://doi.org/10.1002/net.10039}, so these methods are either not scalable or produce large gaps or low quality results. 
%On the large-scale  instances these methods are time costly. 
To ameliorate this issue, we propose a novel algorithm which may leverage Ising processing units (IPUs)~\cite{Coffrin_2019}, specialized computational devices specifically tailored to solve the Ising model, as well as classical solvers that can exhibit high quality solutions for small instances in a reasonable computational time. Examples of IPUs include but are not limited to quantum annealers, gate-based quantum machines equipped with the Ising model  solvers, and both digital and analog annealers~\cite{mohseni2022ising,liu2022leveraging}. Since physical limitations of existing IPU hardware severely restrict the size of problems they can handle and connectivity between variables, we adopt a hybrid IPU-classical approach in which a classical computer decomposes SPNI into subproblems small enough for an IPU, offloads them for computation, then combines the IPU's results into a global solution. For proof of concept, we use exact solvers instead of a real IPU. Our decomposition based results demonstrate 
an ability to achieve almost the same quality as the whole problem slow solvers that are prohibitive for large-scale instances and IPU hardware.
Additionally, our approach is parallelization friendly, namely, sub-problems obtained as a result of SPNI decomposition can be tackled in parallel. 

\section{Related Work}
In one of the earliest works studying SPNI, Malik et al. focused on a variant termed the $k$ most vital arcs problem in which the interdiction of each arc requires exactly one unit of the attacker's budget and interdiction of an arc results in its complete removal \cite{MALIK1989223}. Ball et al. show that the $k$ most vital arcs problem is $\mathcal{NP}$-Hard. Since this is a variant of SPNI, it follows that SPNI is $\mathcal{NP}$-Hard as well \cite{BALL198973}. Corley and Shaw \cite{CORLEY1982157} investigated the single-most-vital-arc problem, which is the $k$ most vital arcs problem where $k = 1$, but this problem is a simple variant of SPNI that lies in $\mathcal{P}$. Rather than study the full removal of arcs, Fulkerson and Harding \cite{10.1007/BF01584329} and Golden \cite{https://doi.org/10.1002/nav.3800250412} study related problems in which the length of an arc increases as more budget is allocated to its interdiction. In one of the most prominent works studying SNPI, Israeli and Wood \cite{https://doi.org/10.1002/net.10039} study the problem in its more general form, proposing two integer-programming based algorithms of different quality depending on whether arcs are fully removed or if arc length is increased as a result of interdiction. 

More recent work on SPNI has since shifted away from integer-programming algorithms. Huang et al., for example, approximate SPNI using a reinforcement learning framework \cite{doi:10.1080/00207543.2021.2002962}. However, this approach generates quite significant gaps and the reinforcement learning is not particularly scalable. %\is{am I right? (Krish) I don't think so, their abstract mentions they run on larger graphs to beat IP methods}. 
Rocco et al. \cite{ROCCOS2010232} study an extension of SPNI where the attacker is not solely focused on maximizing the $s$-$t$ path length, but also wishes to minimize interdiction cost, and propose an evolutionary algorithm to do so. This is a multiobjective problem that is different than ours. %\is{does this approach suffer from something like slowness, large memory? is the objective actually different than ours? (Krish) yes, objective is different as it is multiobjective, unsure about memory as paper does not mention}. 
For more details, we refer the reader to the recent comprehensive survey on the network interdiction problem  \cite{smith2020survey}. To our knowledge, however, no attempts have been made for parallelization friendly decomposition of  SPNI for IPU hardware.

Several frameworks have been proposed to solve large instances of combinatorial optimization problems on small IPU hardware. The quadratic unconstrained binary optimization (QUBO) formulation is a popular formulation for combinatorial optimization problems that is equivalent to the Ising model,  
%\is{DONE add reference, perhaps Glover's survey}
and thus is solvable on all IPU hardware \cite{DBLP:journals/corr/abs-1811-11538}. QUBO formulations contain only quadratic binary variables with no constraints. A large number of work in the quantum computing space focuses on formulating problems into one large QUBO, then identifying sub-QUBOs that can be solved directly on an IPU. The qbsolv tool developed by D-Wave accomplishes exactly this, once allowing researchers to solve a 1254 binary variable problem using quantum hardware that can only solve problems with at most 64 binary variables \cite{10.3389/fict.2017.00029}. An alternative approach is, instead of first formulating the problem as a QUBO then decomposing the QUBO itself, to decompose the problem and then formulate these subproblems into QUBOs. For example, Shaydulin et al. \cite{https://doi.org/10.1002/qute.201900029} take this approach to solve the community detection problem on quantum hardware. This is the approach we opt to take in this paper, applied to SPNI.

\section{Problem Formulation}
Let $G = (N, A)$ be a directed network with node set $N$ and arc set $A$. The length of arc $k \in A$ is $c_k \geq 0$, and the added interdiction length is $d_k \geq 0$, meaning that if arc $k$ is interdicted, then its effective length becomes $c_k + d_k$. If interdiction destroys arc $k$, then $d_k$ can be set to a sufficiently large value. For any node $i \in N$, let $A^+(i) \subseteq A$ denote the set of all arcs directed out of node $i$, and similarly let $A^-(i) \subseteq A$ denote the set of all arcs directed into node $i$. While other forms of SNPI allow for multiple resource types and variant interdiction costs, for our algorithmic specializations we model a single resource type and the scenario in which interdiction of any arc in $A$ costs only one unit of this resource. Thus, let $r_0$ denote the amount of budget we have available. Let $\mathbf{x} = \{x_k\}_{k=1}^{|A|}$ where for each arc $k \in A$, $x_k$ is a binary variable denoting whether arc $k$ is interdicted or not by the attacker, and let $\mathbf{y}= \{y_k\}_{k=1}^{|A|}$ where $y_k$ is a binary variable denoting whether arc $k$ is traversed by the defender. Finally, let $s, t \in N$ denote the source and sink node in $G$, respectively. The integer programming formulation of SNPI is
\begin{equation*}
\begin{array}{ll@{}ll}
\displaystyle \max_{\mathbf{x} \in X} \min_{\mathbf{y}} & \displaystyle \sum_{k \in A} (c_k + x_k d_k) y_k \\
\text{s.t.} & \displaystyle \sum_{k \in A^+(s)} y_{k} - \sum_{k \in A^-(s)} y_{k} = 1 \\
& \displaystyle\sum_{k \in A^+(t)} y_{k} - \sum_{k \in A^-(t)} y_{k} = -1 \\
& \displaystyle\sum_{k \in A^+(i)} y_{k} - \sum_{k \in A^-(i)} y_{k} = 0, \forall i \in N \setminus \{ s, t \},\\
\end{array}
\end{equation*}
where $X = \{ \mathbf{x} \in \{0, 1\}^{|A|} ~\mid ~ \sum_{i=1}^{|A|} \mathbf{x}_i \leq r_0 \}$. \\ \\
Note that the above formulation is biobjective, and thus can not be directly converted into a QUBO model. As Israeli et al. \cite{https://doi.org/10.1002/net.10039} show, we may fix $\mathbf{x}$, take the dual of the inner minimization, make some modifications, then release $\mathbf{x}$, and obtain a single objective formulation
\begin{equation}\label{eq:dual}
\begin{array}{ll@{}ll}
\displaystyle \max_{\mathbf{x}, \mathbf{\pi}} & \pi_t \\
\displaystyle \text{s.t.} & \pi_j - \pi_t - d_k x_k \leq c_k, \forall k = (i, j) \in A \\
\displaystyle & \pi_s = 0 \\
\displaystyle & \mathbf{x} \in X.
\end{array}
\end{equation}
We may interpret $\pi_i$ as the post-interdiction shortest-path length from $s$ to $i$. As such, we may impose bounds on $\pi_i \in [0, |N|\max_{k' \in A} \{c_{k'} + d_{k'} \}]$. Since each variable is constrained and we have a single objective, we may now convert each constraint to a penalty function and decompose bounded variables into several binary variables, for example using the mapping proposed by Karimi et al. \cite{Karimi_2019}. Letting $P$ be a sufficiently large positive penalty value, our QUBO formulation is 
\vspace{-0em}
\begin{equation}
\begin{array}{ll@{}ll}
\displaystyle \max_{\mathbf{x}, \mathbf{\pi}, \mathbf{m}, n} & \displaystyle \pi_t - P \sum_{k = (i, j) \in A} (c_{k} - (m_k + \pi_j - \pi_i - d_{k} x_k))^2 \\ & 
\displaystyle - P(\pi_s)^2 - P(r_0 - (n + \sum_{k \in A} x_k))^2 \\
\text{s.t.} & \displaystyle \pi_i \in [0, |N|\max_{k' \in A} \{c_{k'} + d_{k'} \}], \forall i \in N \\
& \displaystyle m_k \in [0, |N|\max_{k' \in A} \{c_{k'} + d_{k'} \} + d_k], \forall k \in A \\
& \displaystyle n \in [0, r_0] \\
& \displaystyle \mathbf{x} \in \{ 0, 1 \}^{|A|}.
\end{array}
\end{equation}
Note that we have included bounded non-binary variables for simplicity. Additionally, we have included variables $\mathbf{m}$ and $n$ to act as slack variables for inequality constraints. In the actual QUBO formulation, all bounded variables and slack variables should be mapped to several binary variables. The given QUBO formulation is directly solvable on an IPU.

We now describe how to formulate subproblems of SPNI for the purposes of our algorithm. We define a connected partition of a network $G = (N, A)$ as a subset of nodes $N$ such that its induced subgraph is weakly connected.
Say we have a connected partition  of $G$, i.e., a subset of nodes $N_p \subseteq N$, a sink node $t' \in N_p$, and let $A_p$ represent all arcs associated with this subset of nodes, i.e. $A_p = \{ (u, v) \in A | v \in N_p \}$. Note that we may partition $A_p = A_{p,1} \cup A_{p,2}$ where $A_{p,1} = \{ (u, v) \in A | u \in N_p \land v \in N_p \}$ and $A_{p,2} = \{ (u, v) \in A | u \notin N_p \land v \in N_p \}$. We define a subproblem of SPNI over this partition $N_p$ and sink node $t'$ by restricting our free variables to only those associated with $N_p$---rather than define $\pi_i$ for all $i \in N$, we will only define $\pi_i$ for all $i \in N_p$, and similarly only define $x_k$ for all $k \in A_p$. Additionally, we change the objective formula to maximize $\pi_{t'}$ rather than $\pi_t$.

For those variables not associated with $N_p$, we may fix their values to constants. Assume that we have a current solution $A' \subseteq A$---by ``solution,'' we refer to a subset of arcs chosen for interdiction. We use $\Gamma_i$ to denote a constant representing the shortest-path length from node $s$ to $i$ after all arcs in $A'$ have been interdicted. Since $\pi_i$ represents the postinterdiction shortest-path length from nodes $s$ to $i$, we can fix $\pi_i$ to the value $\Gamma_i$ for all nodes $i$ not in $N_p$. Additionally, we limit our budget from $r_0$ to $r_0 - |A' \setminus A_p|$, ensuring we use only the budget we have allocated within $N_p$. The subproblem formulation over $N_p$ with sink node $t'$ is thus

\begin{equation}\label{eq:form}
\begin{array}{ll@{}ll}
\displaystyle \max_{\mathbf{x}, \mathbf{\pi}, \mathbf{m}, n} & \displaystyle \pi_{t'} - P \sum_{k = (i, j) \in A_{p,1}} (c_{k} - (m_k + \pi_j - \pi_i - d_{k} x_k))^2 \\ & 
\displaystyle - P \sum_{k=(i,j) \in A_{p,2}} (c_{k} - (m_k + \pi_j - \Gamma_i - d_{k} x_k))^2 \\ &
\displaystyle - P(\pi_s)^2 \\ & \displaystyle - P((r_0 - |A' \setminus A_p|) - (n + \sum_{k \in A_p} x_k))^2 \\
\text{s.t.} & \displaystyle \pi_i \in [0, |N|\max_{k' \in A} \{c_{k'} + d_{k'} \}], \forall i \in N_p \\
& \displaystyle m_k \in [0, |N|\max_{k' \in A} \{c_{k'} + d_{k'} \} + d_k], \forall k \in A_p \\
& \displaystyle n \in [0, r_0].
\end{array}
\end{equation}
Note that in the above formulation, we have restricted our free variables to those associated with $N_p$, thereby decreasing computational load and potentially allowing the problem to be solved on an IPU depending on the size of $N_p$.

\section{Algorithm}
In this section we present an algorithm for approximating SPNI by leveraging small IPU hardware. Our algorithm first proceeds by generating an initial greedy solution, then refining this solution over several iterations. 

First we define terminology and helper functions. A problem instance of SPNI is a tuple $\mathcal{P} = (G, \mathbf{c}, \mathbf{d}, s, t, r_0)$ where $G = (N, A)$ is a directed network, $\mathbf{c}$ is a vector representing arc lengths of $A$, $\mathbf{d}$ is a vector representing added interdiction arc lengths of $A$, $s \in N$ is the source node, $t$ is the sink node, and $r_0$ is the interdiction budget. 

We define PARTITION to be  a function that takes in a graph $G = (N, A)$ and an integer $n$ as inputs, then returns $P \subseteq \mathcal{P}(V)$, a connected partitioning  of $G$ where $\forall N_p \in P$, $|N_p| \approx n$. These partitions are how we will choose to define subproblems. The purpose of the size restriction is to ensure that subproblems are sufficiently small for IPU hardware. Additionally, PARTITION is necessarily non-deterministic ensuring different partitions on each call to prevent the solver from being stuck in a local attraction basin.

We define CALC-LENGTH, a function that takes in a problem instance $\mathcal{P} = (G = (N, A), \mathbf{c}, \mathbf{d}, s, t, r_0)$, and a set $A' \subseteq A$ where $|A'| \leq r_0$. It returns the shortest $s$-$t$ path length in $G$ after all arcs in $A'$ have been interdicted. 

We also define CALC-PATH that takes in a problem instance $\mathcal{P} = (G = (N, A), \mathbf{c}, \mathbf{d}, s, t, r_0)$, and a set $A' \subseteq A$ where $|A'| \leq r_0$. It returns the set $N' \subseteq N$ consisting of all nodes in the shortest $s$-$t$ path in $G$ after all arcs in $A'$ have been interdicted. 

Finally, function SOLVE takes in a problem instance $\mathcal{P} = (G = (N, A), \mathbf{c}, \mathbf{d}, s, t, r_0)$, a set $N_p \subseteq N$ representing a continuous partition of $G$, a node $t' \in N_p$ for the sink node of the subproblem, and a set of arcs $A' \subseteq A$ where $|A'| \leq r_0$ representing the current working solution. It formulates a subproblem based off these parameters according to \eqref{eq:form}, solves the subproblem, and returns a new arc set $A'' \subseteq A$ representing which arcs to interdict to maximize $\pi_{t'}$. 
%\jl{[DONE $\pi_{t'}$?]}. 
Here is where computational work may be offloaded to an IPU. Note that due to the subproblem restriction, SOLVE can only modify the interdiction status of arcs associated with $N_p$, and thus all arcs not associated with $N_p$ but in $A'$ will remain in $A''$.

\textbf{Initial solution generation} The pseudocode is given in Algorithm~\ref{alg:init}. It receives in input a problem instance $\mathcal{P} = (G = (N, A), \mathbf{c}, \mathbf{d}, s, t, r_0)$ alongside an integer $n$ representing the approximate nodes allowed per partition. To generate an initial solution to SPNI, the algorithm takes a greedy approach. As each arc costs exactly one unit of the budget to interdict, the algorithm iterates over each unit of the budget attempting to decide which arc to allocate this unit to. To make this decision, the algorithm first partitions the network using PARTITION and additionally calculates the shortest $s$-$t$ path with CALC-PATH. Note that it is sufficient to only consider edges along the $s$-$t$ path for interdiction, since edges outside of this path will not affect the $s$-$t$ path length after interdiction. For each node in the $s$-$t$ path, the algorithm determines which partition this node belongs to, then solves a subpartition using this node as the sink node by calling SOLVE. It then takes the output solution, and calls CALC-LENGTH to determine this solution's impact on the $s$-$t$ path length. If this path length is greater than the current maximum path length, the algorithm notes it as a potential candidate solution and updates the maximum. After repeating this process for each node in the $s$-$t$ path, the algorithm chooses a candidate solution randomly, then continues this process until the budget has ran out. Note that in each iteration, the algorithm incrementally adds one arc to the current solution.

\textbf{Solution refinement} The pseudocode is given in Algorithm~\ref{alg:refine}. It receives in input a problem instance $\mathcal{P} = (G = (N, A), \mathbf{c}, \mathbf{d}, s, t, r_0)$, an integer $n$ representing the approximate number of nodes per partition, an integer $\lambda$ for the number of iterations to run the algorithm, and a solution $A' \subseteq A$ to improve. Note that the given solution $A'$ is expected to be the output of Algorithm~\ref{alg:init}. This algorithm attempts to improve the initial solution. To do so, it iterates $\lambda$ times, and in each iteration, partitions the graph using PARTITION and calculates nodes in the current $s$-$t$ path given the current interdiction solution using CALC-PATH. For each of these nodes, alongside all nodes found in the shortest path for the previous iteration, it determines what partition the node lies in and solves a subproblem where this node is the target node using SOLVE. CALC-LENGTH is then called to determine whether the outputted solution's $s$-$t$ path length is greater than the current $s$-$t$ path length, and if so, marks this solution as a potential candidate solution. After iterating over each node, if a better $s$-$t$ path length is found, the current solution is chosen to be a randomly selected candidate solution. If a better $s$-$t$ path length is not found, the algorithm attempts to explore alternate solutions by iterating over each arc in the current solution, temporarily deleting that arc from the solution, and observing whether this deletion can result in a better solution through a partitioning process similar to the above. If a better solution is found, the deletion of this arc becomes permanent and the better solution is adopted. Note that to prevent excessive computation, the algorithm keeps track of arcs which, upon being temporarily deleted, have not previously increased the $s$-$t$ path length in a set named good-arcs. In future iterations, the algorithm does not consider temporarily deleting arcs found in good-arcs.

\begin{algorithm}
\caption{Initial solution generation}
\label{alg:init}
    \SetAlgoLined
    \Require{A problem instance $\mathcal{P} = (G = (N, A), \mathbf{c}, \mathbf{d}, s, t, r_0)$}
    \Require{Approximate number of nodes per partition $n$}
    $A' \leftarrow \{ \}$

    prev-nodes $ \leftarrow $ CALC-PATH$(\mathcal{P}, A')$
    
    $r' \longleftarrow 1$
    
    \While {$r' \leq r_0$}{
    $P \leftarrow \text{PARTITION}(G, n)$

    curr-nodes $ \leftarrow $ CALC-PATH$(\mathcal{P}, A')$

    next-sols $\leftarrow \{A'\}$

    best-length $\leftarrow$ CALC-LENGTH$(\mathcal{P}, A')$

    \ForEach{node $\in$ curr-nodes $\cup$ prev-nodes}{
        $N_p \leftarrow \emptyset$

        \ForEach{$N_p' \in P$}{
            \If{node $\in N_p'$}{
            $N_p \leftarrow N_p'$
            }
        }

        curr-sol $\leftarrow $ SOLVE$(\mathcal{P}, N_p, \text{node}, A')$

        curr-length $\leftarrow$ CALC-LENGTH$(\mathcal{P}, \text{curr-sol})$

        \If{curr-length $>$ best-length}{
        next-sols $\leftarrow \emptyset$
        }
        \If{curr-length $==$ best-length}{
        next-sols $\leftarrow$ next-sols $\cup \{ \text{curr-sol} \}$
        }
    }

    $A' \stackrel{R}{\leftarrow}$ next-sols \Comment{Pick a solution randomly}

    $r' \leftarrow r' + 1$
    }
    \Return $A'$
\end{algorithm}

\begin{algorithm}
\label{alg:refine}
\caption{Solution refinement}\label{euclid}
    \SetAlgoLined
    \Require{A problem instance $\mathcal{P} = (G = (N, A), \mathbf{c}, \mathbf{d}, s, t, r_0)$}
    \Require{Approximate number of nodes per partition $n$}
    \Require{Number of iterations $\lambda$}
    \Require{Current solution $A'$}
    prev-nodes $\leftarrow $ CALC-PATH$(\mathcal{P}, A')$

    next-sols $\leftarrow \{ A' \}$
    
    best-length $\leftarrow $ CALC-LENGTH$(\mathcal{P}, A')$

    good-arcs $\leftarrow \emptyset$
    
    $i \leftarrow 0$
    
    \While{$i < \lambda$}{
    $P \leftarrow $ PARTITION$(G, n)$
    
    curr-nodes $\leftarrow$ CALC-PATH$(\mathcal{P}, A')$

    \ForEach{node $\in$ curr-nodes $\cup$ prev-nodes}{
        $N_p \leftarrow \emptyset$

        \ForEach{$N_p' \in P$}{
            \If{node $\in N_p'$}{
            $N_p \leftarrow N_p'$
            }
        }

        curr-sol $\leftarrow $ SOLVE$(\mathcal{P}, N_p, \text{node}, A')$

        curr-length $\leftarrow$ CALC-LENGTH$(\mathcal{P}, \text{curr-sol})$

        \If{curr-length $>$ best-length}{
        next-sols $\leftarrow \emptyset$
        }
        \If{curr-length $==$ best-length}{
        next-sols $\leftarrow$ next-sols $\cup \{ \text{curr-sol} \}$
        }
    }
    prev-length $\leftarrow$ CALC-LENGTH$(\mathcal{P}, A')$
    
    \uIf{best-length $>$ prev-length}{
    $A' \stackrel{R}{\leftarrow}$ next-sols \Comment{Pick a solution randomly}
    }\uElse{
    \ForEach{arc $\in A' \setminus $ good-arcs}{
    $A'' \leftarrow A' \setminus \{ \text{arc}\}$
    
    \ForEach{node $\in$ curr-nodes $\cup$ prev-nodes}{
        partition = $\emptyset$

        \ForEach{$p \in P$}{
            \If{node $\in p$}{
            partition $\leftarrow p$
            }
        }

        curr-sol $\leftarrow $ SOLVE$(\mathcal{P}, N_p, \text{node}, A'')$

        curr-length $\leftarrow$ CALC-LENGTH$(\mathcal{P}, \text{curr-sol})$

        \uIf{curr-length $>$ prev-length}{
        $A' \leftarrow$ curr-sol
        } \Else{
        good-arcs $\leftarrow$ good-arcs $\cup \{ $ arc $\}$
        }
    }
    }
    }
    }
    \Return A'
\end{algorithm}

\section{Computational Results} 
\begin{figure}[ht]
\centerline{\includegraphics[scale=0.4]{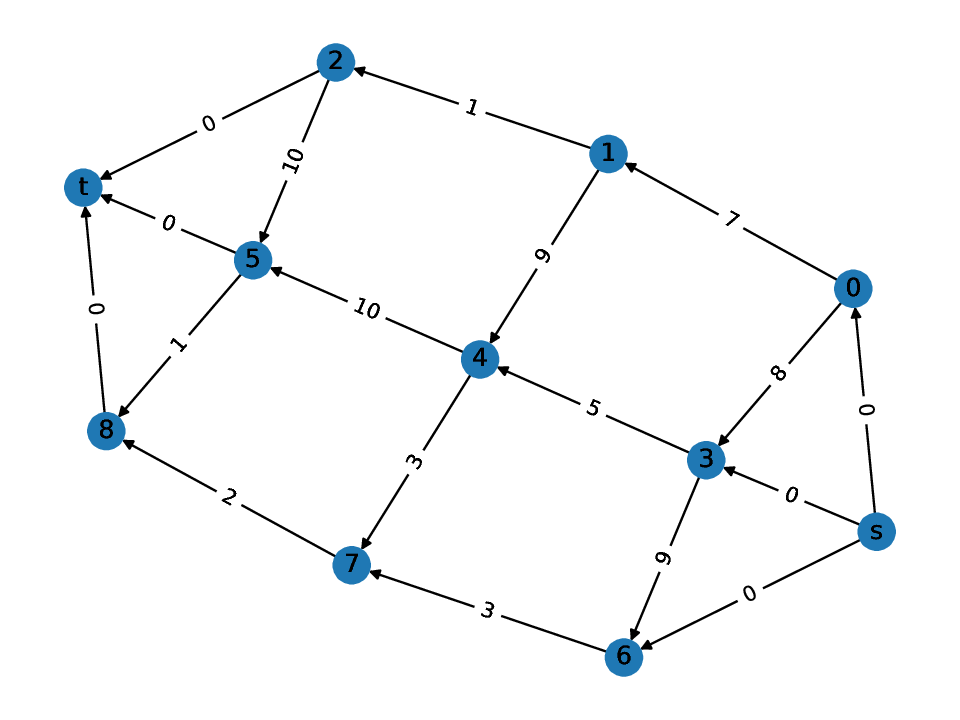}}
\caption{Example of a generated 3x3 grid}
\label{fig}
\end{figure}

To evaluate the performance of our algorithm and similar to other network interdiction literature, we have generated grid networks with randomized edge weights and randomized interdiction weights. To each grid we added the source node $s$  connected to all nodes in the first column, and the sink node $t$  connected to all nodes in the last column, with all of these arcs having length $0$ and interdiction length $0$ to make the problem more challenging. All horizontal arcs are oriented forward and all vertical arcs are oriented down. All arcs not connected to $s$ or $t$ have a randomly assigned length and interdiction length, each in the range $[1, 10]$. Figure~\ref{fig} depicts an example of grid 3x3.

Models were built using Pyomo, and were solved using the open-source solver CBC  \cite{forrest2005cbc}. No significant difference was observed in experiments with Gurobi solver \cite{gurobi}. %from COIN-OR (www.coin-or.org).
Graph representation and utilities such as Dijkstra's was done using NetworkX \cite{hagberg2008exploring}. Graph partitioning was done using METIS  \cite{Karypis_1998}. All problems were ran on the University of Delaware's Caviness cluster which uses a Linux system and has exclusively Intel E5-2695 v4 18 core processors in its pool.

% The ``$s$-$t$ Path Length'' column refers to the shortest $s$-$t$ path length without any interdicted arcs. The ``Refinement Path Length'' and ``Refinement Time (s)'' columns refers to the shortest $s$-$t$ path length post interdiction found by the refinement algorithm, and how long the algorithm took to find this solution in CPU time, respectively. The ``Classical Path Length'' column refers to the shortest $s$-$t$ path length post interdiction found by taking the original dual formulation on the entire graph and solving it directly---note that we only run the classical solver for as much CPU time as the refinement solver took.

Below we plot the results of our computational experiments. The ``Size'' axis represents the number of nodes in the experiment's network ($\vert N\vert$). Note that by the structure of the grid network, the number of arcs can be calculated by $\vert A \vert = 2(\vert N\vert - 2)$. The ``Quality'' axis displays a metric to evaluate the quality of our solutions (defined in Eq. (\ref{eq:qual})). 

We run both our refinement algorithm and CBC attempting to solve Equation~\eqref{eq:dual} across the entire network---we shall hereby refer to the latter as the ``full problem solver.'' Note that the full problem solver is significantly slower than the refinement algorithm, and thus, we timeout the full problem solver once it has exceeded the time that the refinement algorithm takes, then proceed to take the current best solution. \emph{In every problem instance below, the full problem solver always timed out}. For any given experiment, if $r$ is the shortest $s$-$t$ path length after interdicting the arcs from the solution of our refinement algorithm and $f$ the shortest $s$-$t$ path length after interdicting the arcs from the full problem solver's solution, the ratio can be calculated by 
\begin{equation}\label{eq:qual}
(r - f)/\max(r, f).
\end{equation}
Thus, the higher the quality, the better---positive quality indicates that the refinement solver performed better than the full problem solver.

\begin{figure}[htbp]
\centerline{\includegraphics[scale=0.4]{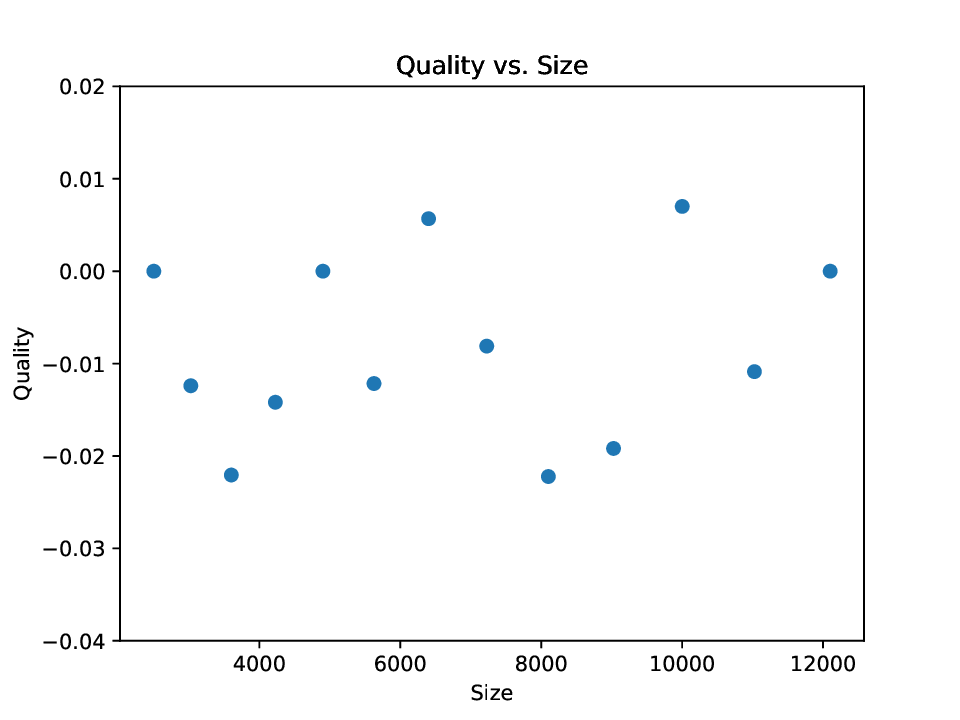}}
\caption{Computational results for different size networks with parameters: $\lambda = 50$, $n=20$, $r_0 = 0.0025\vert A \vert$. Each point corresponds to a graph. The horizontal axis corresponds to the number of nodes. The vertical axis is the quality defined in Eq. (\ref{eq:qual}).}
\label{fig:225}
\end{figure}

\begin{figure}[htbp]
\centerline{\includegraphics[scale=0.4]{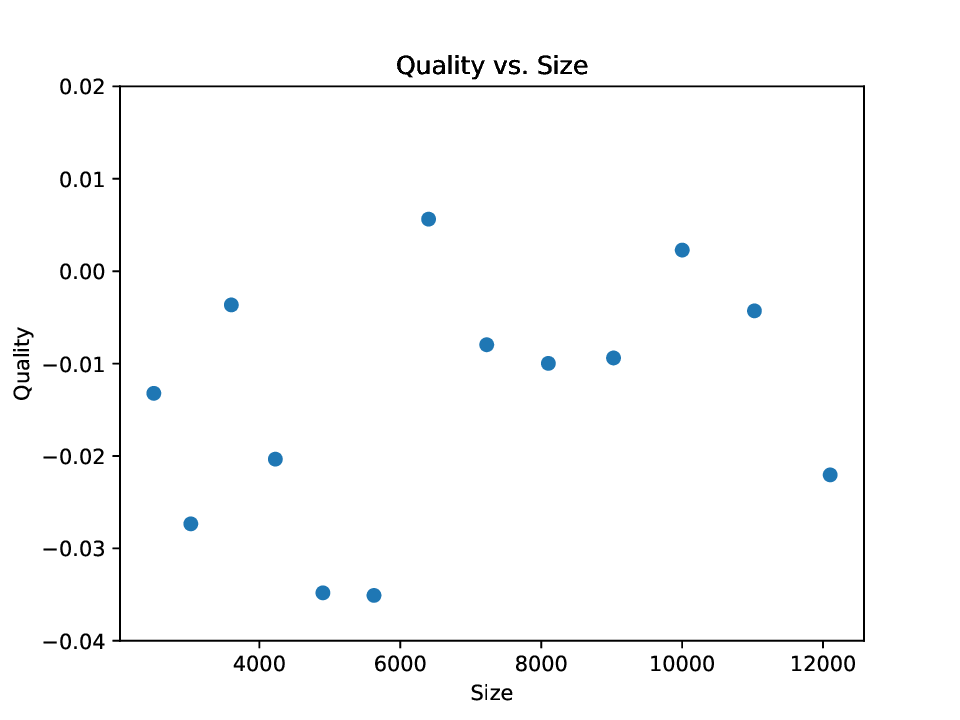}}
\caption{Computational results for different size networks with parameters: $\lambda = 50$, $n=20$, $r_0 = 0.05\vert A \vert$. Each point corresponds to a graph. The horizontal axis corresponds to the number of nodes. The vertical axis is the quality defined in Eq. (\ref{eq:qual}).}
\label{fig:250}
\end{figure}

We run three sets of experiments. In each experiment, we set the  number of refinement iterations $\lambda$ to 50. (However, the convergence is often observed earlier.) In the first set of experiments, represented by Figure~\ref{fig:225}, we set the interdiction budget $r_0$ to $0.25\%$ of the number of arcs in the network and limit our subproblem size to $n=20$ nodes. The average quality for this set of experiments is $-0.008$, implying that a subproblem size of only 20 nodes is sufficient for a good estimation of SPNI. In the second set of experiments, represented by Figure~\ref{fig:250}, we set the interdiction budget $r_0$ to $0.5\%$ of the number of arcs in the network and limit our subproblem size to $n=20$ nodes. The average quality for this set of experiments is $-0.013$, indicating relative consistency in estimation regardless of budget. In the final set of experiments, we set the interdiction budget $r_0$ to $0.25\%$ of the number of arcs in the network and limit our subproblem size to $n = 40$ nodes. The average quality for this set of experiments is $-0.012$, implying that \emph{an increase in subproblem size is not necessary for estimation and limited capacity IPU machines can already be integrated to solve this problem using decomposition-based refinement}.

Out of all 39 experiments ran, the difference between the $s$-$t$ path length produced by the refinement solver and the path length produced by the full problem solver was $\geq -10$ for 35 of the experiments. In 10 of these experiments, the refinement solver performed better than the full problem solver. \emph{These results are significant as they indicate that in the cases when full problem solver was better the refinement missed the value of a single arc.}%, since arc lengths are randomly selected from the range [1, 10]}.

\begin{figure}[htbp]
\centerline{\includegraphics[scale=0.4]{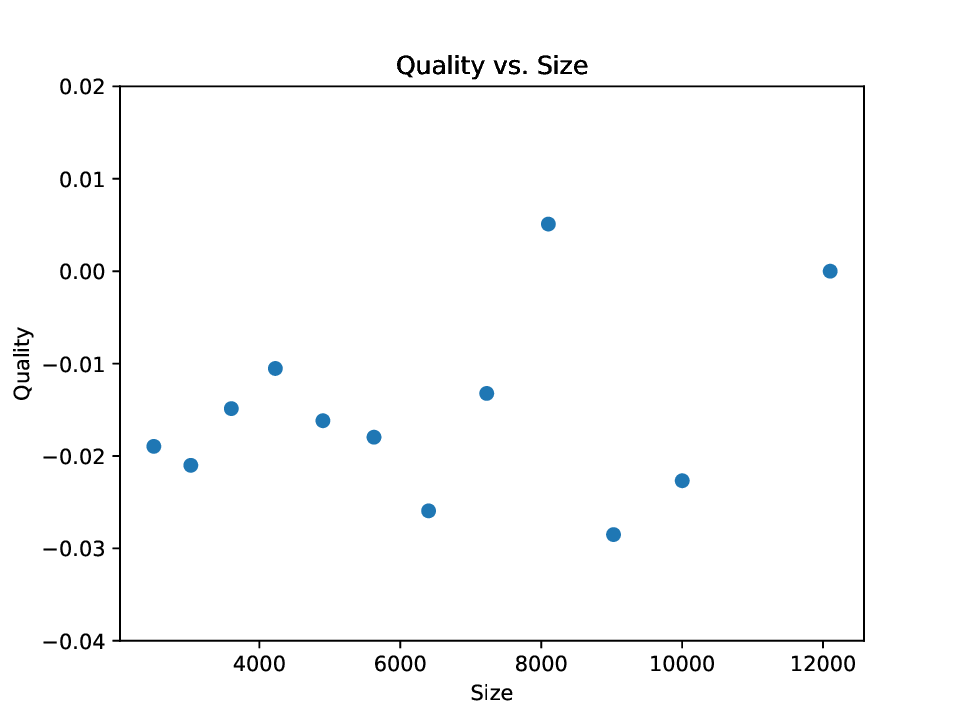}}
\caption{Computational results for different size networks with parameters: $\lambda = 50$, $n=40$, $r_0 = 0.0025\vert A \vert$. Each point corresponds to a graph. The horizontal axis corresponds to the number of nodes. The vertical axis is the quality defined in Eq. (\ref{eq:qual}).}
\label{fig:425}
\end{figure}

\section{Discussion}
\subsection{Solving on Ising Processing Hardware}
The current results shown in this paper rely on using an exact solver to solve subproblems. We have previously attempted to utilize classical \emph{heuristic} QUBO solvers, such as D-Wave's \emph{qbsolv} and D-Wave Ocean, to solve these subproblems, but such solvers provided poor quality solutions even on much smaller problem sizes that those we experiment with. %(e.g. unable to find positive solutions on networks with only 10 nodes, resulting in 77 free binary variables).
Additionally, due to lack of accessibility to massive experiments with the IPU hardware, utilizing exact solvers is necessary to demonstrate the efficiency of our algorithm. We hope that as IPUs continue to develop in capacity, their superior solution quality and speed when compared to classical heuristic solvers will be able to make full use of the algorithm presented in this paper, generating solutions that are able to beat classical exact solvers consistently.

% Here you need to analyze the total number of variables for a problem to solve on IPU. Mention that we have tried QUBO heuristics but they were not good at it, so having exact solvers is important. In other words, make the case for IPU.
\subsection{Algorithmic Challenges and Obstacles}
Refinement algorithms for solving SPNI may face several challenges, the most notable being the problem of proper budget allocation amongst partitions. If an algorithm decides to improperly allocate budget to a subproblem, for example deciding to interdict an arc where this unit of budget would ideally be spent much further away in the graph, it proves to be very difficult to shift this budget significantly far away from its current position. Within our refinement approach, the movement of interdicted arcs is primarily dictated by random partitioning---for example, in one iteration, the subproblem solver may determine that a specific arc should be interdicted under the context of the current partititioning configuration, i.e. within the context of the current subproblem the arc lies in. But when we repartition in the next iteration, the subproblem solver reevaluates the placement of interdicted arcs---the cost unit used to interdict the previous arc may be relocated elsewhere based on the new subproblem formulation. Since the cost unit used for an interdicted arc may only shift to arcs relatively near this arc, it is unlikely for the cost unit to travel far away in the graph. To resolve this issue, a multilevel approach widely known in combinatorial scientific computing and applied graph algorithms  \cite{inproceedings} may be taken. Multilevel algorithms create a hierarchy of coarse problem representations, find a solution to the coarsest (smallest and most compressed)  problem, then gradually derive a solution to the original finest  problem by projecting the solution back from the coarse to fine levels and refining it using our algorithm, thereby potentially aiding in proper budget allocation. This approach has seen great success in solving various (hyper)graph problems \cite{1639360,safro2011multiscale,shaydulin2019relaxation} including hybrid quantum-classical multi-  and single-level refinement for partitioning and community detection \cite{https://doi.org/10.1002/qute.201900029,ushijima2021multilevel}, indicating that its application to SPNI may improve solution quality.

\subsection{Performance}
One of our main goals was to develop a trivial IPU  parallelization friendly algorithm. Our current implementation of the proposed refinement  does not take advantage of the fact that subproblems produced by a given partitioning can be solved independently of one another, since they only depend on the current working solution.  
Consequently, subproblems can be solved in parallel rather than sequentially, offering a potentially large optimization to the current performance of the algorithm.

\subsection{Generalizability}
The algorithm introduced in this paper does not handle SPNI problem instances in which the binary interdiction of a given arc may cost more than one unit of the interdictor's budget. The algorithm can currently handle instances where an arc between two nodes can take on multiple values of interdiction (e.g., if interdicted once add $1$ to the length, if interdicted twice add $2$ to the length) through the introduction of multi-edges with appropriate lengths, but instances in which a binary decision needs to be made (i.e., interdicted or not) and differing levels of interdiction cost are not supported.

Another scenario which the algorithm is unable to handle is in the case of different resource types. For example of such a scenario, let $r_0$ and $r_1$ represent the budget for two different types of resources, then arc $k_0$ may cost $1$ of $r_0$ and $2$ of $r_1$, but arc $k_1$ may cost $0$ of $r_0$ and $3$ of $r_1$. This restriction is not much of a concern, however, considering that in one of the most prominent works regarding SPNI,  the authors of \cite{https://doi.org/10.1002/net.10039} also do not accommodate this case for their algorithmic cases.

\section{Conclusion}
Solving SPNI holds crucial importance in the defense, engineering and other domains, addressing challenges in securing critical  infrastructure from terrorist attacks, natural disasters and disrupting enemy supply networks among other applications. Current solutions to SPNI, however, are often too slow to be scalable to large networks, and thus prove to be impractical for real world purposes. In this paper, we have introduced a novel decomposition approach addressing SPNI harnessing the rapid solving abilities of IPU hardware to yield an approximate solution. We have further shown that solutions generated by our decomposition algorithm are extremely close in quality to state-of-the-art integer programming methods. We anticipate that as IPU hardware advancements continue to improve speed in solving optimization problems, our algorithm will enable an even quicker approximation of SPNI thereby facilitating a more efficient approach for real world scenarios of SPNI. More work is under way to produce higher quality approximations in even shorter amounts of time and extend the algorithm to more general forms of SPNI.\\
\noindent {\bf Reproducibility: } Our results and code are available at \url{https://github.com/krishxmatta/network-interdiction}.

\section*{Acknowledgements}
We would like to thank J. Cole Smith and Yongjia Song, the authors of \cite{smith2020survey}, for their input and discussion about the current status of the network interdiction solvers.

This work was supported  in part with funding from the Defense Advanced Research Projects Agency (DARPA). The views, opinions and/or  findings expressed are those of the author and should not be interpreted as representing the official views or policies  of the Department of Defense or the U.S.Government.

This research was supported in part through the use of DARWIN computing system: DARWIN - A Resource for Computational and Data-intensive Research at the University of Delaware and in the Delaware Region, which is supported by NSF Grant \#1919839.

\bibliographystyle{plain}
\bibliography{main}
\end{document}